\documentclass[%
 reprint,
nofootinbib,
 amsmath,amssymb,
 aps,
]{revtex4-2}

\usepackage{graphicx}
\usepackage{float}
\usepackage{dcolumn}
\usepackage{amsfonts} 
\usepackage{dsfont}
\usepackage{amssymb}
\usepackage{svg}
\usepackage{amsmath}
\usepackage{hyperref}
\usepackage{subcaption}
\usepackage{slashed}
\usepackage{mathtools}

\begin{document}

\preprint{APS/123-QED}

\title{Quantum Fisher information of the Klein--Gordon and Dirac vacua}

\author{Preslav Asenov}
\email{preslav.asenov.20@ucl.ac.uk}
\author{Alessio Serafini}
\email{serale@theory.phys.ucl.ac.uk}

\affiliation{Department of Physics and Astronomy, University College London,
 Gower Street, London WC1E 6BT, United Kingdom}

\date{\today}

\begin{abstract}
The quantum Fisher information (QFI) of the vacuum of three quantum field theories is evaluated with respect to the mass parameter of each theory. All field theories are considered on a $(d+1)$-dimensional Euclidean spacetime. We consider the Klein-Gordon field and the free Dirac field. For the free KG case, we find a $m^{d-2}$ dependence of the QFI, and thus no dependence for $d=2$, in agreement with the holographic duality characterizing the theory. The vacuum QFI with respect to the free Dirac field mass is shown to be UV-divergent for $d=2$ and $d=3$, mass-dependent for $d=1$, and zero for $d=0$.

\end{abstract}

\maketitle
\section{Introduction}


The geometrization of physics is a long-standing methodological doctrine spanning a range of research areas \cite{kalinowski1988program}. Following the geometrization of gravitation in general relativity \cite{einstein1916grundlagen}, the philosophy of seeking geometric reformulations of physical phenomena has been carried over to classical mechanics \cite{ghaboussi1993geometrization,arnold1989mathematical}, quantum mechanics \cite{kibble1979geometrization}, and quantum field theory \cite{gomes1945gauge,alonso2024geometric}. Geometric quantum mechanics, in particular, reformulates quantum mechanics by representing the physical characteristics of a quantum system as geometric structures on the system’s quantum state space \cite{brody2001geometric}. Since, as we shall see, this connects naturally to the estimation of quantum mechanical parameters, it is worthwhile to give a short overview of how this is done. First notice that, in quantum mechanics, a physical system is associated with a complex Hilbert space $\mathcal{H}$. 
Here, we consider an infinite-dimensional $\mathcal{H}$ with vectors $|\psi(\boldsymbol{\theta})\rangle \in \mathcal{H}$ that depend on the $m$-dimensional continuous parameter $\boldsymbol{\theta} = ( \theta_1, \theta_2, \dots, \theta_m) \in \mathcal{M}$, where $\mathcal{M}$ is an $m$-dimensional manifold. A physical state of the system under consideration is represented by the ray $[|\psi(\boldsymbol{\theta})\rangle]$, defined by
\begin{equation}
[|\psi(\boldsymbol{\theta})\rangle] = \{ |\phi(\boldsymbol{\theta})\rangle \in \mathcal{H}^{\times} : |\phi(\boldsymbol{\theta})\rangle = c |\psi(\boldsymbol{\theta})\rangle, \, c \in \mathbb{C}^{\times} \} ,
\end{equation}
where we adopt the notation $V^{\times} = V \setminus \{0\}$ for any vector space $V$. Note that the following treatment is restricted to pure states only (see Ref. \cite{gibbons1992typical} for a treatment including mixed states). The infinite-dimensional, complex projective space $P(\mathcal{H})$ of $\mathcal{H}$ is defined as the set of all rays as \cite{brody1998statistical}
\begin{equation}
P(\mathcal{H}) = \{ [|\psi(\boldsymbol{\theta})\rangle] : |\psi(\boldsymbol{\theta})\rangle \in \mathcal{H}^{\times} \} .
\end{equation}
$P(\mathcal{H})$ can also be viewed as a real manifold \cite{minic2003background} equipped with an integrable almost-complex structure, a symplectic form, and a Riemannian metric (the Fubini-Study metric \cite{fubini1904sulle,study1905kurzeste}). Hence, $P(\mathcal{H})$ is a K\"ahler manifold \cite{chruscinski2004geometric} \footnote{More specifically, $P(\mathcal{H})$ is a special kind of K\"ahler manifold with a constant holomorphic sectional curvature. See Ref. \cite{ashtekar1999geometrical} for more details.}. The map
\begin{align}
    \Psi : \quad & \mathcal{M} \to \mathcal{H} \nonumber \\[5pt]
    & \boldsymbol{\theta} \mapsto |\Psi(\boldsymbol{\theta})\rangle
\end{align}
can be used to induce the pullback of the Fubini-Study metric on $P(\mathcal{H})$ to a Riemannian metric on $\mathcal{M}$ \cite{venuti2007quantum}, called the quantum information metric (QIM). The QIM is given by $g^{(\Psi)}_{ab} = \text{Re}[G^{(\Psi)}_{ab}]$, where $G^{(\Psi)}_{ab}$ is the quantum geometric tensor (QGT) on $\mathcal{M}$ given by \cite{provost1980riemannian}
\begin{align}
    & G^{(\Psi)}_{ab} (\boldsymbol{\theta}) = \langle \partial_a \Psi(\boldsymbol{\theta}) | \partial_b \Psi(\boldsymbol{\theta}) \rangle \nonumber \\[5pt]
    & - \langle \partial_a \Psi(\boldsymbol{\theta}) | \Psi(\boldsymbol{\theta}) \rangle \langle \Psi(\boldsymbol{\theta}) | \partial_b \Psi(\boldsymbol{\theta}) \rangle.
\end{align}
Here we use the shorthand $|\partial_a \Psi(\boldsymbol{\theta}) \rangle = \partial_a |\Psi(\boldsymbol{\theta}) \rangle$, where $\partial_a = \frac{\partial}{\partial \theta_a}$ is the partial derivative with respect to the component $\theta_a$ of $\boldsymbol{\theta}$. Note that $\omega^{(\Psi)}_{ab} = \text{Im} \big[G^{(\Psi)}_{ab}\big]$ is the Berry curvature \cite{berry1984quantal,berry1984quantal,aharonov1987phase}. The QGT (the main object of investigation in quantum geometry) has been used to study quantum phase transitions \cite{zanardi2007information,gu2010fidelity}, to optimize variational quantum circuits \cite{stokes2020quanutm}, and has found numerous other applications in condensed matter physics (see Refs. \cite{gao2025quantum,yu2025quantum} for reviews on the subject). 

The QGT feature that we consider in the following is its application in quantum metrology, or quantum estimation theory (QET) -- the framework that considers the problem of estimating one or multiple parameters on which a quantum state may depend, based on a measurement of the given state \cite{helstrom1969quantum, paris2009quantum}. QET offers methods to establish the precision limits constraining such measurements, as well as, in principle, schemes that achieve these limits and potentially surpass the performance of classical systems \cite{giovannetti2011advances}. A key concept in QET is the quantum Fisher information matrix $I^{(\psi)}_{ab}$ (QFIM) of a quantum state $|\psi(\boldsymbol{\theta})\rangle$, calculated with respect to the parameters $\boldsymbol{\theta}$ to be estimated \cite{paris2009quantum}. The QFIM allows one to place a fundamental precision bound on parameter estimation based on quantum measurements, known as the quantum Cram\'{e}r-Rao lower bound (QCRLB). QFIM calculations have been applied in a variety of contexts, including condensed matter physics \cite{dellanna2023quantum, mazza2024quantumfisherinformationstrange}, cosmology \cite{gomez2021quantum, chen2024quantum, piotrak2025quantum}, and optics \cite{zhang2012unbounded, tsang2019resolving, hradil2019quantum}. For pure states, $I^{(\psi)}_{ab} = 4g^{(\psi)}_{ab} = 4 \text{Re}\big[G^{(\psi)}_{ab}\big]$. 

The QGT and the QFIM have also brought insight in the context of the anti-de Sitter/conformal field theory (AdS/CFT) correspondence \cite{ammon2015gauge}. 
Miyaji et al. \cite{miyaji2015distance} showed that the vacuum-state quantum Fisher information (QFI) with respect to the coupling constant of a CFT is approximately dual to the volume of a codimension one time slice in AdS space. This work presented a method for calculating the QGT \footnote{As correctly pointed out by Ref. \cite{dimov2021holographic}, the terminology established in Refs. \cite{miyaji2015distance,trivella2017holographic}, with a slight abuse of notation, refers to $G_{ab}$ as the quantum information metric, instead of the quantum geometric tensor. Note that we only consider real, single-parameter $G_{aa}$, so the two objects are equivalent here.}, and hence the QFIM, of the vacuum state of a conformal field theory (CFT) with respect to the coupling constants of the theory (Note that, while critiques of this proposal exist \cite{moosa2018volume,belin2019complexity}, the validity of the QFIM calculation technique used here has not been put into question). Refs. \cite{bak2016information, trivella2017holographic,banerjee2018connecting,erdmenger2020information,dimov2021holographic} then introduced further developments to this work. As stated in Ref. \cite{miyaji2015distance}, this calculation technique is general to any quantum field theory (QFT) and is not specific to CFTs. Ref. \cite{alvarez2017quantum} applied this method to different quantum systems and pointed out that its application to any QFT is a possible continuation of this line of investigation. To the best of our knowledge, up to now no work has used this or any other method to find the QGT of a non-conformal QFT state with respect to the parameter(s) of the QFT Lagrangian. The aim of the present work is to address this gap. We use the method proposed by Miyaji et al. to consider a QFT in a flat, $(d+1)$-dimensional Euclidean spacetime background (i.e. in imaginary time) and to then calculate the QFI of the QFT vacuum state with respect to a parameter of the QFT Lagrangian. Namely, we consider the mass parameter for a free Klein-Gordon (KG) field and a free Dirac field.

This calculation places our work in the context of a wider effort to bridge quantum information and QFT \cite{erdmenger2020information,faulkner2022snowmass}. More specifically, methods from QET, including QFIM calculations, have been applied to optimize the estimation of parameters, such as relative distances, gravitational field strengths, proper times and accelerations when measuring pure probe states of a QFT \cite{ahmadi2014relativistic,ahmadi2014quantum,lindkvist2015motion}. These methods have been extended to the context of measuring mixed states \cite{vsafranek2015ultimate} and have been used to place a precision bound on the detection of gravitational waves that are shown to create phonons in a Bose–Einstein condensate \cite{sabin2014phonon}. QET methods have also been applied in the context of the Unruh effect \cite{aspachs2010optimal,wang2014quantum,yao2014quantum,tian2015relativistic,feng2022quantum}.
Ref. \cite{junior2017geometry} calculated the classical Fisher information matrix (CFIM) of the probability distribution associated with measuring a particular field configuration in a massless KG QFT. The notion of proximity between quantum field theories, as opposed to quantum states, was investigated in Ref. \cite{balasubramanian2015relative} The CFIM in Ref. \cite{junior2017geometry} was calculated with respect to the coordinate labels of a flat spacetime background. Calculations of the classical and quantum Fisher information with respect to the kinematic parameters of two-particle states in QED were performed in Ref. \cite{asenov2026quantum}.  Furthermore, Ref. \cite{ignoti2025large} calculated the quantum and classical Fisher information with respect to the Dirac CP-violating phase in the context of neutrino oscillation detection. Finally, we note that there exist alternative QFI calculation approaches to the one proposed by Miyaji et al. The first computation of the QFI of the vacuum state of a CFT was performed in Ref. \cite{lashkari2016canonical} using the relation of QFI to the second order variation of relative entropy between two states. A real-time path integral representation of the QFI of a QFT state is also given in Ref. \cite{headley2026path} (see also Ref. \cite{ilias2022critcality}).

This paper is structured as follows. Section \ref{subsec:FI} introduces the concept of (classical and quantum) Fisher information in the context of (classical and quantum) estimation theory. Section \ref{subsec:vacuum_qfi_of_qft} reviews the method proposed by Miyaji et al. (as presented in Ref. \cite{trivella2017holographic}). Section \ref{sec:method_results} presents the method and results of our vacuum-state QFI calculation with respect to the mass of a KG (\ref{subsec:kg_qim}) and a free Dirac (\ref{subsec:free_dirac_qim}) field theory. Finally, Section \ref{sec:conclusion} gives a conclusion and a discussion of our results.

\subsection{Fisher Information} \label{subsec:FI}

\subsubsection{Classical Fisher Information}
Here we introduce the concept of classical Fisher information (CFI) in classical estimation theory (CET). Consider a classical probability distribution with a joint probability density function (PDF) $p(\textbf{y}) = p(\textbf{y}|\boldsymbol{\theta})$, describing the probability that a vector of random variables $\textbf{Y} = ( Y_1, Y_2, \dots, Y_n)$ \cite{devore2011probability} is physically observed to take the value $\textbf{y} = (y_1, y_2, \dots, y_n)$ after a single measurement of a classical system. The PDF considered here depends on a vector parameter $\boldsymbol{\theta} = ( \theta_1, \theta_2, \dots, \theta_m)$. The task of CET is to find an estimator $\hat{\boldsymbol{\theta}}(\textbf{y})$  (a map from the space of observations $\textbf{y}$ to the parameter manifold $\mathcal{M}$ \cite{zielinski1997theory}) that is close to the true vector parameter value $\boldsymbol{\theta}$ \cite{helstrom1969quantum}. A lower bound can be placed on the variance $\text{Var} [ \hat{\boldsymbol{\theta}}(\textbf{y}) ]$ -- this is given by the classical  Cram\'{e}r-Rao lower bound (CCRLB), which depends on the classical Fisher information matrix $I^{(C)}_{\boldsymbol{\theta},\textbf{y}}$ (CFIM) of $p(\textbf{y}|\boldsymbol{\theta})$ with respect to the vector parameter $\boldsymbol{\theta}$. The CFIM is a positive semi-definite, symmetric $m \times m$ matrix, with its $(a,b)$ element given by \cite{ly2017tutorial}
\begin{equation}
    \left( I^{(C)}_{\boldsymbol{\theta},\textbf{y}} \right)_{ab} =
    \text{Cov} \left[L^{(C)}_a , L^{(C)}_{b} \right].
\end{equation}
Here $L^{(C)}_a = \frac{\partial \ \text{log} \ p(\textbf{y}|\boldsymbol{\theta})}{\partial \theta_a}$ is defined as the score function, with its derivative evaluated at the true value of $\theta_a$ \cite{pickles1985introduction,kay1993fundamentals}, while the covariance $\text{Cov} \left[ X,Y \right]$ of two random variables $X$ and $Y$ is given by $\text{Cov} \left[ X,Y \right] = \text{E}\left[ \left( X - \text{E}\left[ X \right] \right) \left( Y - \text{E}\left[ Y \right] \right) \right]$. Then, for an unbiased estimator with expectation value $\text{E} \left[ \hat{\boldsymbol{\theta}}(\textbf{y}) \right] = \boldsymbol{\theta}$ \cite{devore2011probability} the CCRLB
\cite{kay1993fundamentals} is given by
\begin{equation}
    \text{Cov} \left[\hat{\boldsymbol{\theta}}(\textbf{y})\right] \geq \frac{1}{N} \left[I^{(C)}_{\boldsymbol{\theta},\textbf{y}}\right]^{-1},
\end{equation}
where $\big[I^{(C)}_{\boldsymbol{\theta},\textbf{y}}\big]^{-1}$ is the inverse of the CFIM and $N$ is the number of independent $\textbf{Y}$ measurements performed on $p(\textbf{y}|\boldsymbol{\theta})$. The CCRLB sets the ultimate limit to the achievable precision achievable when estimating a parameter based on a measurement of a PDF.

In the single-parameter estimation case, where $\boldsymbol{\theta} = \theta$, $I^{(C)}_{\boldsymbol{\theta},\textbf{y}}$, the CFIM reduces to the classical Fisher information (CFI) $I^{(C)}_{\theta,\textbf{y}}$ given by \cite{wasserman2004all}
\begin{align} \label{fisher_info_definition1}
    I^{(C)}_{\theta,\textbf{y}} &= \text{Var} \left[ L^{(C)}_{\theta,\textbf{y}} \right] 
     = \text{E} \left[ \left( L^{(C)}_{\theta,\textbf{y}}\right)^2 \right],
\end{align}
where $L^{(C)}_{\theta,\textbf{y}} = \frac{\partial\, \text{log}\, p(\textbf{y}|\theta)}{\partial \theta}$. Then, the single-parameter CCRLB reduces to
\cite{kay1993fundamentals}
\begin{equation}
    \text{Var} \left[\hat{\theta} (\textbf{y})\right] \geq \frac{1}{N I^{(C)}_{\theta,\textbf{y}}}.
\end{equation}

\subsubsection{Quantum Fisher Information}
The following description of QET is adapted from \cite{paris2009quantum}. In QET, we describe our system by the density operator $\rho (\boldsymbol{\theta})$ on $\mathcal{H}$ rather than the PDF $p(\textbf{y}|\boldsymbol{\theta})$ \cite{helstrom1969quantum}. The two are related by the Born rule $p(\textbf{y}|\boldsymbol{\theta}) d\textbf{y} = \text{Tr} \left[ d\Pi_{\textbf{y}} \rho (\boldsymbol{\theta}) \right]$, where $d\Pi_{\textbf{y}}$ are positive operator-valued measure (POVM) elements satisfying $\int d\Pi_{\textbf{y}} = {\mathds{1}}$. In QET, the (classical) score function $L^{(C)}_a$ is generalized to the symmetric logarithmic derivative (SLD) quantum operator $L_a$. Since in the CET scenario $L^{(C)}_a p(\textbf{y}|\boldsymbol{\theta}) = \partial_{\theta_a}{p(\textbf{y}|\boldsymbol{\theta})}$ is satisfied, in QET the SLD is defined by 
\begin{equation}\label{sld_definition}
    \frac{1}{2} \left( L_a \rho (\boldsymbol{\theta}) + \rho (\boldsymbol{\theta}) L_a \right) = \partial_a \rho (\boldsymbol{\theta}).
\end{equation}
Then, the quantum Fisher information matrix (QFIM) is a positive semi-definite, symmetric $m \times m$ matrix, with its $(a,b)$ element given by
\begin{equation} \label{eq:Fisher_multiparam_mixed}
    \left( I^{(\rho)}_{\boldsymbol{\theta}} \right)_{ab} = \text{Tr} \left[ \rho (\boldsymbol{\theta}) \frac{L_{\theta_a} L_{\theta_b} + L_{\theta_b} L_{\theta_a}}{2} \right].
\end{equation}
For a pure state $\rho (\boldsymbol{\theta}) = \rho^2 (\boldsymbol{\theta}) = |\psi ({\boldsymbol{\theta}}) \rangle \langle \psi ({\boldsymbol{\theta}})|$ Eq. (\ref{eq:Fisher_multiparam_mixed}) reduces to \cite{liu2019quantum}
\begin{align} \label{eq:Fisher_multiparam_single}
     \left( I^{(\psi)}_{\boldsymbol{\theta}} \right)_{ab} = &4\text{Re}\big[ \langle \partial_a \psi ({\boldsymbol{\theta}}) |\partial_b \psi ({\boldsymbol{\theta}}) \rangle \nonumber \\[5pt]
    & - \langle \partial_a \psi ({\boldsymbol{\theta}}) | \psi ({\boldsymbol{\theta}}) \rangle \langle \psi ({\boldsymbol{\theta}}) | \partial_b \psi ({\boldsymbol{\theta}}) \rangle \big] .
\end{align}
For $N$ independent quantum measurements of $\textbf{Y}$ on $\rho (\boldsymbol{\theta})$, the covariance of any estimator $\hat{\boldsymbol{\theta}}$ is constrained by the QCRLB bound as:
\begin{equation}\label{eq:qcrlb_general}
    \text{Cov} \left[ \hat{\boldsymbol{\theta}}(\textbf{y}) \right] \geq \frac{1}{N} \bigg[I^{(\psi)}_{\boldsymbol{\theta}}\bigg]^{-1},
\end{equation}
where $\big[I^{(\psi)}_{\boldsymbol{\theta}}\big]^{-1}$ is the inverse of the QFIM.

In the single-parameter estimation scenario $\boldsymbol{\theta} = \theta$, the QFIM $I^{(\psi)}_{\boldsymbol{\theta}}$ reduces to the quantum Fisher information (QFI)
\begin{equation}\label{quantum_fisher_info_definition1}
I^{(\psi)}_{\theta} = \text{Tr} \left[ \rho (\theta) L_{\theta}^2 \right].
\end{equation}
The CCRLB bounds the variance of the single-parameter estimator as
\begin{equation}\label{eq:single_var_qcrlb}
    \text{Var}\left[\hat{\theta}(\textbf{x})\right] \geq \frac{1}{N I^{(\psi)}_{\theta}}.
\end{equation}
Note that once a POVM (i.e., a specific quantum measurement) is set, the Born rule extracts a classical probability distribution from a quantum state, and one can apply the classical analysis presented above, whereby the variance of a parameter $\theta$ is minimized by the (achievable) CCRLB. The optimization of the CCRLB over all possible POVMs then gives the QCRLB in terms of the QFI. The latter is therefore just the classical Fisher information optimized over all possible POVMs. Hence, the QFIM maximizes the CFIM over all possible POVMs as
\begin{equation}
    I^{(\psi)}_{\boldsymbol{\theta}} \geq I^{(C)}_{\boldsymbol{\theta},\textbf{y}}
\end{equation}
and the QCRLB is a lower bound of the CCRLB.

As an aside, we note that for some differentiable, invertible, smooth function $f(\theta)$ of the parameter $\theta$ with a non-zero derivative $\partial_{\theta}f(\theta)$, the pure-state SLD $L_{f(\theta)}$ with respect to $f(\theta)$ is related to $L_{\theta}$ by $L_{\theta} = \frac{\partial f(\theta)}{\partial \theta} L_{f(\theta)}$ if $\rho (f(\theta))=\rho (\theta)$. Then, the QFI with respect to $f(\theta)$ is
\begin{align} \label{eq:relation_i_f_theta_to_i}
I^{(\psi)}_{\theta} = \left( \frac{\partial f(\theta)}{\partial \theta}\right)^2 I^{(\psi)}_{f(\theta)}.
\end{align}

\subsection{Vacuum-state quantum Fisher information of a quantum field} \label{subsec:vacuum_qfi_of_qft}

Here we present a short review of the general method developed to find the QGT of any quantum field, as shown in Ref. \cite{alvarez2017quantum}.
In this approach a quantum field $\varphi(x) = \varphi (\textbf{x},\tau)$ with Lagrangian density $\mathcal{L}_0$ is considered in a $(d+1)$-dimensional Euclidean spacetime with coordinates $x = (\textbf{x},\tau)$, divided into spatial coordinates $\textbf{x} = ( \, \text{x}^{i} \, | \,  i=1,\dots,d \, )$ and Euclidean time $\tau$. The quantum field is described by $\mathcal{L}_0$ for $\tau \in (-\infty,0]$ and has a ground state $|\Omega_0\rangle$. At $\tau=0$ a perturbation is introduced to $\mathcal{L}_0$, giving the perturbed Lagrangian density 
\begin{align}
    \mathcal{L}_1 &= \mathcal{L}_0+\delta\mathcal{L} 
    = \mathcal{L}_0+\delta \boldsymbol{\theta} \cdot \boldsymbol{\mathcal{O}}. 
\end{align}
Here $\boldsymbol{\mathcal{O}} = (\mathcal{O}_1, \dots, \mathcal{O}_m)$ are perturbation operators and $\boldsymbol{\theta}$ is a vector of physical field parameters. $\delta\boldsymbol{\theta}$ is a very small deformation in these physical parameters under which $\boldsymbol{\theta} \rightarrow \boldsymbol{\theta} + \delta\boldsymbol{\theta}$. $\mathcal{L}_1$ describes the quantum field in the range $\tau \in [0,\infty)$ and has a ground state $|\Omega_1\rangle$. In order to compute the QGT of this quantum field, we are interested in computing the fidelity $\mathcal{F}(\boldsymbol{\theta},\boldsymbol{\theta} + \delta\boldsymbol{\theta})$ between the vacuum states $|\Omega_0\rangle$ of $\mathcal{L}_0$ and $|\Omega_1\rangle$ of $\mathcal{L}_1$. We consider $|\Omega_1\rangle$ in the limit $\tau \rightarrow -\infty$ and $|\Omega_0\rangle$ in the limit $\tau \rightarrow +\infty$. Then, we define the fidelity between the two pure ground states as
\begin{align} \label{eq:fidelity_definition}
    \mathcal{F}(\boldsymbol{\theta},\boldsymbol{\theta} + \delta\boldsymbol{\theta}) &\equiv |\langle \Omega_1|\Omega_0\rangle| \nonumber \\[5pt]
    & = 1 - \frac12 \sum^n_{a,b=1} G_{ab} \, \delta \theta_a \, \delta \theta_b + O((\delta \boldsymbol{\theta})^3), 
\end{align}
where $\langle \Omega_1|\Omega_0\rangle$ is shorthand for $\langle\Omega_0(\tau \rightarrow -\infty, \boldsymbol{\theta})|\Omega_1(\tau \rightarrow +\infty, \boldsymbol{\theta}+ \delta\boldsymbol{\theta})\rangle$. In order to compute $\mathcal{F}(\boldsymbol{\theta},\boldsymbol{\theta} + \delta\boldsymbol{\theta})$, the overlap between $|\Omega_0\rangle$ and $|\Omega_1\rangle$ is expressed as the path integral
\begin{equation} \label{eq:vacuum_overlap_1}
    \langle \Omega_1|\Omega_0\rangle = \int_{-\infty}^{\infty}\mathcal{D} \tilde{\varphi} \, \langle \Omega_1|\tilde{\varphi}\rangle \langle \tilde{\varphi}|\Omega_0\rangle.
 \end{equation}
Here $|\tilde{\varphi}\rangle = |\tilde{\varphi}_{\tau=0}(\vec{x})\rangle$ is an eigenstate of the field operator $\tilde{\varphi}(x)=\varphi(\vec{x},0)$ at $\tau = 0$ in the Schr\"odinger picture. Since $|\Omega_0\rangle$ is defined for $\tau\in(-\infty,0]$ and $|\tilde{\varphi}\rangle$ is defined at $\tau=0$, the overlap $\langle \tilde{\varphi}|\Omega_0\rangle$ is given by the path integral
\begin{equation}
    \langle \tilde{\varphi}|\Omega_0\rangle = \frac{1}{\sqrt{Z_0}} \int_{\tilde{\varphi}} \mathcal{D}\varphi \, \text{exp}\bigg( -\int_{-\infty}^{0} d\tau \int_{-\infty}^{\infty} d^d\textbf{x} \, \mathcal{L}_0 \bigg),
\end{equation}
where the Euclidean evolution from $\tau=-\infty$ to $\tau=0$ is considered, and the functional integral is taken over fields $\tilde{\varphi}(\textbf{x},\tau) = \varphi(x,0)$ . Here $Z_0$ is the partition function corresponding to $\mathcal{L}_0$ and is given by
\begin{equation}
    Z_0 = \int_{-\infty}^{\infty}\mathcal{D}\varphi \, \text{exp}\bigg( -\int_{-\infty}^{\infty} d\tau \int_{-\infty}^{\infty} d^d\textbf{x} \, \mathcal{L}_0 \bigg).
\end{equation}
Similarly, the overlap between $|\tilde{\varphi}\rangle$ and $|\Omega_1\rangle$, which is defined for $\tau\in[0,\infty)$, considers the Euclidean evolution from $\tau=0$ to $\tau=+\infty$ and is hence given by the path integral
\begin{align}
    & \langle \Omega_1 | \tilde{\varphi} \rangle \nonumber \\[5pt]
    & = \frac{1}{\sqrt{Z_1}} \int_{\tilde{\varphi}} \mathcal{D}\varphi \, \text{exp}\bigg( -\int_{0}^{\infty} d\tau \int_{-\infty}^{\infty} d^d\textbf{x} (\mathcal{L}_0 + \delta\boldsymbol{\theta} \cdot \boldsymbol{\mathcal{O}})\bigg),
\end{align}
where 
\begin{equation}
    Z_1 = \int_{-\infty}^{\infty}\mathcal{D}\varphi \, \text{exp}\bigg( -\int_{-\infty}^{\infty} d\tau \int_{-\infty}^{\infty} d^d\textbf{x} (\mathcal{L}_0 + \delta\boldsymbol{\theta} \cdot \boldsymbol{\mathcal{O}}) \bigg).
\end{equation}
These expressions are used to evaluate Eq. (\ref{eq:vacuum_overlap_1}):
\begin{align}
    & \langle \Omega_1 | \Omega_0 \rangle = \frac{1}{\sqrt{Z_0 Z_1}} \int_{-\infty}^{\infty}\mathcal{D}\varphi\, \text{exp} \bigg( -\int_{-\infty}^{0} d\tau \int_{-\infty}^{\infty} d^d\textbf{x} \mathcal{L}_0 \nonumber \\[5pt]
    &-\int_{0}^{\infty} d\tau \int_{-\infty}^{\infty} d^d\textbf{x} (\mathcal{L}_0 + \delta\boldsymbol{\theta} \cdot \boldsymbol{\mathcal{O}}) \bigg),
\end{align}
which can be written as
\begin{equation} \label{eq:vacuum_overlap_zero_lim}
    \langle \Omega_1 | \Omega_0 \rangle = \frac{\langle \Omega_0 | \text{exp} \big( -\int_{0}^{\infty} d\tau \int_{-\infty}^{\infty} d^d\textbf{x} \,\delta \boldsymbol{\theta} \cdot \boldsymbol{\mathcal{O}} \big) |\Omega_0 \rangle }{\sqrt{\langle \Omega_0 | \text{exp} \big( -\int_{-\infty}^{\infty} d\tau \int_{-\infty}^{\infty} d^d\textbf{x} \,\delta \boldsymbol{\theta} \cdot \boldsymbol{\mathcal{O}} \big)|\Omega_0 \rangle}}.
\end{equation}
Expanding Eq. \ref{eq:vacuum_overlap_zero_lim} in $\delta \boldsymbol{\theta}$ and comparing the second-order term to Eq. \ref{eq:fidelity_definition} gives the QGT expression
\begin{align} \label{eq:general_qim}
     G_{ab, d+1} = &
    \int^{0}_{-\infty} d\tau \int_{0}^{\infty}d\tau' \int_{-\infty}^{\infty} d^d \textbf{x} \int_{-\infty}^{\infty} d^d \textbf{x}' \nonumber \\[5pt]
    & \cdot \bigg( \langle \Omega_0 | \mathcal{O}_a(x) \mathcal{O}_b(x') |\Omega_0\rangle \nonumber \\[5pt] 
    & - \langle \Omega_0 | \mathcal{O}_a(x)|\Omega_0\rangle  \langle \Omega_0 |\mathcal{O}_b(x') |\Omega_0\rangle \bigg).
\end{align}

\section{Method and results}\label{sec:method_results}

In this section we calculate the QGT $G_{m,d+1}$, and hence the QFI $I_{m,d+1}$, of the vacuum state $|0\rangle$ (we revert to a standard notation for the vacuum here, since we do not need to distinguish between different vacua for different values of the parameters) with respect to the mass parameter $m$ (for a QFT on a $(d+1)$-dimensional Euclidean spacetime). Hence, we place the ultimate precision limit (the QCRLB) on any estimation of $m$ that could be in principle obtained by measuring  $|0\rangle$ and, most importantly, establish the analytical dependence of the QFI on the physical parameters. We do this calculation for the mass of a KG and a free Dirac field theory. We use the notation $\langle \cdot \rangle$ and $\langle \cdot \rangle_{D}$ to denote the expectation value over the vacuum state of the KG and the free Dirac field theory, respectively. The same notation $m$ is used throughout, where it is implicit that we denote the mass of the respective field for each chapter. 

Note that we define the QGT and QFI in the context of quantum states in a Hilbert space, while we compute the QFI of the vacuum within the framework of Euclidean QFT. In Euclidean QFT we consider integration measures over classical fields rather than operator-valued distributions acting explicitly on a Hilbert space \cite{zinn2021quantum}. However, correlation functions in Minkowski background QFT (i.e. Wightman distributions) \cite{wightman1956quantum} can be analytically continued to correlation functions in Euclidean QFT (i.e. Schwinger functions) \cite{schwinger1958euclidean} that satisfy the Osterwalder–Schrader axioms \cite{osterwalder1973axioms}. All QFI calculations in the following rely on the correlation functions of the KG and free Dirac field theories in Euclidean spacetime. The vacuum of a QFT in Euclidean spacetime is also well-defined \cite{zinn2021quantum}.

\subsection{Klein-Gordon field theory} \label{subsec:kg_qim}
The KG Lagrangian in $(d+1)$-dimensional Euclidean spacetime is given by \cite{peskin1995introduction}
\begin{equation}
    \mathcal{L}^{(KG)}_0 = \frac{1}{2} \partial_{\mu} \phi \partial_{\mu} \phi + \frac{m^2}{2}\phi^2 ,
\end{equation}
where $\partial_\mu = (\partial_\tau, \nabla)$. We now apply the mass perturbation 
\begin{equation}
\delta \mathcal{L}^{(KG)} = \delta(m^2) \frac{\phi^2}{2} 
\end{equation}
to $\mathcal{L}_0$ such that
\begin{equation}
    \mathcal{L}^{(KG)}_1 = \mathcal{L}^{(KG)}_0 + \delta \mathcal{L}^{(KG)} = \frac{1}{2} \partial_{\mu} \phi \partial^{\mu} \phi + \frac{m^2+\delta{(m^2)}}{2}\phi^2,
\end{equation}
Therefore, we are considering a single-parameter estimation scenario where the deformation parameter is $\theta=m^2$ and the deformation operator is $\mathcal{O}(x)=\frac{\phi(x)^2}{2} $. Substituting this into Eq. \ref{eq:general_qim} gives the QGT on the KG vacuum with respect to the KG mass as
\begin{align} \label{eq:kg_qim_mass_1}
    G^{(KG)}_{m^2,d+1} =& \frac{1}{2^2} \int_{-\infty}^{0}d\tau \int^{\infty}_{0}d\tau' \int_{-\infty}^{\infty} d^d \textbf{x} \int_{-\infty}^{\infty} d^d \textbf{x}' \nonumber \\[5pt]
    & \cdot \bigg( \langle \phi^2(x) \phi^2(x') \rangle - \langle \phi^2(x) \rangle \langle 
    \phi^2(x') \rangle \bigg).
\end{align}
By applying Wick's theorem\cite{wick1950evaluation} we can simplify the integrand as
\begin{align}
    &\langle \phi^2(x) \phi^2(x') \rangle - \langle \phi^2(x) \rangle \langle 
    \phi^2(x') \rangle \nonumber = 2 \big[\Delta(x'-x)\big]^2.
\end{align}
Here, $\Delta(x'-x) = \Delta(x-x')$ is the two-point correlation function of the KG field in $(d+1)$-dimensional Euclidean spacetime given by \cite{curiel2020summing}
\begin{align} \label{eq:kg_propagator_def}
    \Delta(x'-x) & = \int_{-\infty}^{\infty}\frac{d^{d+1}p}{(2\pi)^{d+1}} \frac{e^{-ip\cdot(x'-x)}}{p^2 + m^2} \nonumber \\[5pt]
    & = \frac{m^{\frac{d-1}{2}}}{(2 \pi)^{\frac{d+1}{2}}}|x'-x|^{\frac{1-d}{2}} K_{\frac{d-1}{2}} \big( m|x'-x| \big) ,
\end{align}
where $K_{\alpha} (\cdot)$ is the modified Bessel function of the second kind of order $\alpha$. A property that is used in this paper is
\begin{equation} \label{eq:bessel_symmetry}
    K_{-\alpha} (\cdot) = K_{\alpha} (\cdot) .
\end{equation}
Note that $G^{(KG)}_{m^2,d+1} = \text{Re}[G^{(KG)}_{m^2,d+1}]$, so $I^{(KG)}_{m^2,d+1} = 4G^{(KG)}_{m^2,d+1}$. Furthermore, using Eq. (\ref{eq:relation_i_f_theta_to_i}) we arrive at the relation
\begin{align} \label{eq:qim_m2_to_m}
    & I^{(KG)}_{m,d+1} = 4 m^2 \, I^{(KG)}_{m^2,d+1} .
\end{align}
Substituting these results into Eq. (\ref{eq:kg_qim_mass_1}) gives the KG vacuum QFI with respect to the KG mass as
\begin{align} \label{eq:kg_qim_m_first}
     I^{(KG)}_{m,d+1} 
    = & \frac{2^3 m^{d+1}}{ (2 \pi)^{d+1}} \int_{-\infty}^{0}d\tau \int^{\infty}_{0}d\tau' \int_{-\infty}^{\infty} d^d \textbf{x} \int_{-\infty}^{\infty} d^d \textbf{x}' \nonumber \\[5pt]
    & \cdot |x' - x|^{1-d} K^2_{\frac{d-1}{2}} \big( m|x'-x| \big) .
\end{align}
In order to evaluate this expression we use the following result for $d > 1$
\begin{align} \label{eq:I_1_result}
    & \int^{0}_{-\infty} d\tau \int_{0}^{\infty}d\tau' \int_{-\infty}^{\infty} d^d \textbf{x} \int_{-\infty}^{\infty} d^d \textbf{x}' \, F(|x'-x|) \nonumber \\[5pt]
    & = \frac{2 \pi \cdot V_d}{d} \, \prod^{d-1}_{k=2} \bigg( \frac{\sqrt{\pi} \cdot \Gamma \big( \frac{1 + d - k}{2} \big)}{\Gamma \big( \frac{2 + d - k}{2} \big)} \bigg) 
     \int_{0}^{\infty} dR \cdot F (R) \cdot R^{d+1} ,
\end{align}
where $R = |x' - x|$ is the radial coordinate in $(d+1)$-dimensional coordinates. Here $V_d = \int_{-\infty}^{\infty} d^d \textbf{x}$ is the volume of $d$-dimensional space and $\Gamma(\cdot)$ is the gamma function \cite{abramowitz1948handbook}
\begin{equation}
    \Gamma (z) = \int_{0}^{\infty} s^{z-1}e^{-t} \, ds,\ \qquad \text{Re} [z]>0.
\end{equation}
In the case where $d=1$, we use the following expression instead:
\begin{align} \label{eq:I_1_result_d_1}
 & \int^{0}_{-\infty} d\tau \int_{0}^{\infty}d\tau' \int_{-\infty}^{\infty} d \textbf{x} \int_{-\infty}^{\infty} d \textbf{x}' \, F(|x'-x|) \nonumber \\[5pt]
 & = 2 V_1 \int_{0}^{\infty} dR \cdot F \big( R \big) \cdot R^{2}.
\end{align}
See Appendix \ref{subsec:KG_integral} for a full derivation of these results. Letting
\begin{align}
    F \big( R \big) = \frac{2^3 m^{d+1}}{(2\pi)^{d+1}} R^{1-d} \, K^2_{\frac{d-1}{2}} \big( m R \big),
\end{align}
we evaluate $I^{(KG)}_{m,d+1}$ for $d>1$ as
\begin{align}
     I^{(KG)}_{m,d+1} = & \frac
    {2^3 V_d \cdot m^{d+1}}{d \cdot (2\pi)^d} \, \prod^{d-1}_{k=2} \bigg( \frac{\sqrt{\pi} \cdot \Gamma \big( \frac{1 + d - k}{2} \big)}{\Gamma \big( \frac{2 + d - k}{2} \big)} \bigg) \nonumber \\[5pt]
    & \cdot \int_{0}^{\infty} dR \cdot R^2 \cdot K^2_{\frac{d-1}{2}} \big( m R \big) 
\end{align}
and for $d=1$ as
\begin{align}
     I^{(KG)}_{m,1+1} = & \frac
    {2^2 V_1 m^2}{\pi^2} \cdot \int_{0}^{\infty} dR \cdot R^2 \cdot K^2_{0} \big( m R \big) .
\end{align}
The final results for $d=1,2,3$ are
\begin{equation} \label{eq:kg_qfi_reults}
    I^{(KG)}_{m,1+1} = \frac{V_1}{2^3 m} , \ \ \ I^{(KG)}_{m,2+1} = \frac{V_2}{2^3 \pi} , \ \ \ I^{(KG)}_{m,3+1} = \frac{mV_3}{2^4 \pi}.
\end{equation}
All QFI quantities calculated in this paper depend on $V_d$. It is due to this term that the vacuum states of the considered QFTs contain an infinite amount of information about the QFT mass. Hence, an optimal estimator $\hat{m}$ based on a measurement of $|0\rangle$ has a variance bounded by $\text{Var}[\hat{m}] \geq 0$ once the limit $V_d \to \infty$ is taken. 
Despite the divergence of the QFI terms we compute, it is informative to examine their dependence on the physical parameter $m$. Eq. (\ref{eq:kg_qfi_reults}) shows a $I^{(KG)}_{m,d+1} \propto m^{d-2}$ dependence for $d=1,2,3$. 
Consider the massless limit $m \to 0$, in which the KG field becomes conformally invariant \cite{mcavity1995conformal,poland2019conformal}. $I^{(KG)}_{m,2+1}$ has no $m$ dependence, hence $I^{(KG)}_{m\to 0,2+1}$ is a well-defined QFI of the vacuum state of a CFT. Following Ref. \cite{miyaji2015distance}, this term is then proposed to be dual to the volume of a codimension one time slice
in AdS space. Also, taking the massless limit leads to a divergence of $I^{(KG)}_{m \to 0,1+1}$ in $m$. 
Finally, in the special case where $d=0$, we use the following result
\begin{align} \label{eq:I_0_d0}
    & \int^{0}_{-\infty} d\tau \int_{0}^{\infty}d\tau' \, F(|\tau'-\tau|) = \int_{0}^{\infty} dR_0 \cdot F \big( R_0 \big) \cdot R_0 ,
\end{align}
where $R_0 = \tau' - \tau$. This relation is derived in Appendix \ref{subsec:KG_integral} (see Eq. (\ref{eq:I_1_result_d_0_appendix})). Hence, we evaluate Eq. (\ref{eq:kg_qim_m_first}) as
\begin{align}
    & I^{(KG)}_{m,0+1} = \frac{1}{2 m^2}.
\end{align}
$(0 + 1)$-dimensional QFT is equivalent to quantum mechanics \cite{zee2010quantum} and the free KG Lagrangian is equivalent to the Lagrangian of a quantum harmonic oscillator (QHO) in one dimension \cite{Srednicki_2007}. It can be shown that the $I^{(KG)}_{m,0+1}$ value calculated here is a special case of the calculation of the QIM (or the quantum metric tensor) of a generalized harmonic oscillator in quantum mechanics done in Ref. \cite{gonzalez2019classical}.


\subsection{Free Dirac field theory}
\label{subsec:free_dirac_qim}
Consider the free Dirac field theory Lagrangian in 
$(d+1)$-dimensional Euclidean spacetime \cite{di2023cp} 
\begin{equation}
    \mathcal{L}^{(D)}_0 = \bar{\psi}(\gamma^{0} \partial^{0} + \vec{\gamma}\cdot \vec{\partial})\psi + m\bar{\psi}\psi,
\end{equation}
where $\bar{\psi} = \psi^{\dagger} \gamma^0$ is the Dirac adjoint of the field $\psi$. We now apply the mass perturbation 
\begin{equation}
\delta \mathcal{L}^{(D)} = \delta m\cdot \bar{\psi}\psi
\end{equation}
to $\mathcal{L}^{(D)}_0$, such that
\begin{equation}
    \mathcal{L}^{(D)}_1 = \bar{\psi}(\gamma^{0} \partial^{0} + \vec{\gamma}\cdot \vec{\partial})\psi + (m+\delta m)\bar{\psi}\psi.
\end{equation}
Then, the QGT of the Dirac field vacuum with respect to the Dirac mass $m$ is
\begin{align} \label{eq:dirac_qim_mass_1}
    &G^{(D)}_{m,d+1} = \int_{-\infty}^{0}d\tau \int^{\infty}_{0}d\tau' \int_{-\infty}^{\infty} d^d \textbf{x} \int_{-\infty}^{\infty} d^d \textbf{x}' \nonumber \\[5pt]
    & \bigg( \langle \bar{\psi}(x) \psi(x) \bar{\psi}(x') \psi(x') \rangle_D - \langle \bar{\psi}(x) \psi(x) \rangle_D \langle
    \bar{\psi}(x') \psi(x') \rangle_D \bigg).
\end{align}
Using Wick's theorem, this expectation value simplifies as
\begin{align}
     & \langle \bar{\psi}(x) \psi(x) \bar{\psi}(x') \psi(x') \rangle_D - \langle \bar{\psi}(x) \psi(x) \rangle_D \langle \bar{\psi}(x') \psi(x') \rangle_D \nonumber \\[5pt]
     & = - \text{Tr} [\Delta_D(x-x')\Delta_D(x'-x) ].
\end{align}
Here we use shorthand notation for the free Dirac field propagator $\big( \Delta_D(x-x')\big)_{\alpha \beta} = \langle \psi_{\alpha}(x) \bar{\psi}_{\beta}(x') \rangle_D$ in $(d+1)$-dimensional Euclidean spacetime. The Dirac indices $\alpha$ and $\beta$, and the trace operator $\text{Tr}[\cdot]$ taken over them, are left implicit in the following. Substituting this result into Eq. (\ref{eq:dirac_qim_mass_1}) gives the QGT as
\begin{align} \label{eq:dirac_qim_mass_2}
    G^{(D)}_{m,d+1} = & -\int_{-\infty}^{0}d\tau \int^{\infty}_{0}d\tau' \int_{-\infty}^{\infty} d^d \textbf{x} \int_{-\infty}^{\infty} d^d \textbf{x}' \nonumber \\[5pt]
    & \Delta_D(x'-x)\Delta_D(x-x').
\end{align}
The Dirac propagator in $(d+1)$-dimensional Euclidean spacetime is \cite{ceyhan2013algebraic}
\begin{align}
     \Delta_D(x'-x) = & \left(\frac{m}{2\pi |x'-x|}\right)^{\frac{d+1}{2}} \cdot \bigg[ |x'-x| K_{\tfrac{d-1}{2}}(m|x'-x|) \nonumber \\[5pt]
    & - \sum_{i=0}^{d}\gamma^i(x_i'-x_i) \, K_{\tfrac{d+1}{2}}(m|x'-x|) \bigg] ,
\end{align}
as derived in Appendix \ref{sec:dirac_propagator}. Substituting $\Delta_D(x'-x)$ into Eq. (\ref{eq:dirac_qim_mass_2}) and using the expression of $I^{(D)}_{m,d+1} = 4 \text{Re}[G^{(D)}_{m,d+1}] = 4G^{(D)}_{m,d+1}$ for a real QGT gives the Dirac vacuum QFI with respect to the Dirac mass $m$ as
\begin{widetext}
\begin{equation} \label{eq:dirac_qim_1}
    I^{(D)}_{m,d+1} =
     4 \left(\frac{m}{2\pi}\right)^{d+1} \int^{0}_{-\infty}d\tau \int_{0}^{\infty}d\tau' \int_{-\infty}^{\infty} d^d \textbf{x} \int_{-\infty}^{\infty} d^d \textbf{x}' \, \frac{1}{|x'-x|^{d-1}} 
    \bigg[ K^2_{\tfrac{d+1}{2}}(m|x'-x|) - K^2_{\tfrac{d-1}{2}}(m|x'-x|) \bigg] ,
\end{equation}
where we used $(\gamma^i)^2=1$ and the Dirac algebra in Euclidean spacetime $\{\gamma^i,\gamma^j\}=2\delta^{ij}$ for $i,j \in \{ 0, \dots, d \}$, to simplify the sum
\begin{equation}
    \bigg( \sum_{i=0}^{d} \gamma^i(x_i'-x_i) \bigg)^2 = \sum_{i=0}^{d} (\gamma^i)^2 \, (x_i'-x_i)^2 
     + \sum_{i=0}^{d} \sum_{j=i+1}^{d} \{\gamma^i,\gamma^j \} \, (x_i'-x_i) \, (x_j'-x_j) \nonumber \\[5pt]
     = \sum_{i=0}^{d} (x_i'-x_i)^2 
     = |x'-x|^2.
\end{equation}
\end{widetext}
Note that Eq. (\ref{eq:dirac_qim_1}) is an integral of the form given in Eq. (\ref{eq:I_1_result}) for $d>1$ and of the form given in Eq. (\ref{eq:I_1_result_d_1}) for $d=1$. We use this result to evaluate $I^{(D)}_{m,d+1}$ for $d>0$ by substituting 
\begin{align}
    & F (R) = 4 \left(\frac{m}{2\pi}\right)^{d+1} \cdot \frac{1}{R^{d-1}} \cdot \bigg[ K^2_{\tfrac{d+1}{2}}(mR) - K^2_{\tfrac{d-1}{2}}(mR) \bigg] 
\end{align}
into Eq. (\ref{eq:I_1_result}) for $d>1$ and Eq. (\ref{eq:I_1_result_d_1}) for $d=1$. This gives
\begin{align} \label{eq:dirac_qfi_final_d}
    I^{(D)}_{m,d+1} = & \left(\frac{m}{2\pi}\right)^{d+1} \frac{2^3 \pi  V_d}{d} \cdot \prod^{d-1}_{k=2} \bigg( \frac{\sqrt{\pi} \Gamma \big( \frac{1 + d - k}{2} \big)}{\Gamma \big( \frac{2 + d - k}{2} \big)} \bigg) \nonumber \\[5pt]
    &  \int_{0}^{\infty} dR \bigg[ K^2_{\tfrac{d+1}{2}}(mR) - K^2_{\tfrac{d-1}{2}}(mR) \bigg] \cdot R^{2} .
\end{align}
and 
\begin{align} 
    I^{(D)}_{m,1+1} = & \frac{2 V_1 m^2}{\pi^2} \cdot \int_{0}^{\infty} dR \bigg[ K^2_{1}(mR) - K^2_{2}(mR) \bigg] \cdot R^{2} .
\end{align}
which simplifies to
\begin{equation}
    I^{(D)}_{m,1+1} = \frac{V_1}{2^3 m} .
\end{equation}
To compute $I^{(D)}_{m,d+1}$ for $d>1$, we introduce a (short-distance) ultra violet (UV) cut-off $k$ to the radial coordinate variable $R = |x' - x|$ in $I^{(D)}_{m,d+1}$ as $\int_{0}^{\infty} dR \to \lim_{k \to 0^+} \int_{k}^{\infty}$. Substituting this into Eq. (\ref{eq:dirac_qfi_final_d}) and evaluating the $R$ integral gives the divergent Dirac vacuum QFI in $d=2$ spatial dimensions as
\begin{align}
    & I^{(D)}_{m,2+1} = \lim_{k \to 0^+} \frac{V_2}{2^2 \pi} \big( 1 - \gamma - \log(2km) \big) .
\end{align}
Here 
$\gamma$ is Euler's constant defined by \cite{abramowitz1948handbook}
\begin{equation}
    \gamma = \lim_{n\rightarrow \infty} \bigg( -\text{ln}(n)+\sum_{k=1}^{n}\frac{1}{k}\bigg) \approx 0.577.
\end{equation}
For $d=3$, the QFI $I^{(D)}_{m,3+1}$ also diverges as
\begin{align}
    I^{(D)}_{m,3+1} = \lim_{k \to 0^+} \left( \frac{2^2 V_3}{3 \pi^2 k} - \frac{3 m V_3}{2^4 \pi} \right).
\end{align}
In the special case where $d=0$, Eq. (\ref{eq:dirac_qim_1}) reduces to
\begin{align} 
    I^{(D)}_{m,0+1} = & 4 \left(\frac{m}{2\pi}\right) \int^{0}_{-\infty}d\tau \int_{0}^{\infty}d\tau' \nonumber \\[5pt]
    & (\tau' - \tau) \cdot \bigg[ K^2_{\frac{1}{2}}(m(\tau' - \tau)) - K^2_{\frac{1}{2}}(m(\tau' - \tau)) \bigg] \nonumber \\[5pt]
     = &0 ,
\end{align}
where we used Eq. (\ref{eq:bessel_symmetry}). This contrasts the non-zero QFI result we obtained in the case of a scalar field theory in $(0+1)$ dimensions. The only non-divergent QFI values with respect to the free Dirac field mass are the ones considering a field on $(0+1)$- and $(1+1)$-dimensional background. Interestingly, $I^{(KG)}_{m,1+1} = I^{(D)}_{m,1+1}$. Hence, the two quantities exhibit the same divergence in the massless theory limit, for which the QCRLB of the mass estimator goes to $\text{Var}[\hat{m}] \geq 0$.

\section{Conclusion} \label{sec:conclusion}
In this work we computed the QFI of the vacuum of three QFTs with respect to the mass parameter of the QFT Lagrangian. For the free KG field, QFI expressions were obtained for $d=0,1,2,3$ spatial dimensions, exhibiting critical behavior in the massless theory limit in $d=0$ and $d=1$, while the $d=2$ QFI for $m=0$ is dual to he volume of a codimension one time slice in AdS space according to the proposal in Ref. \cite{miyaji2015distance}. The QFI quantities for a free Dirac field are shown to be UV-divergent for $d=2$ and $d=3$. The vacuum carries no information about the Dirac mass for $d=0$, while in $d=1$, the QFI quantities for the free Dirac and KG fields exhibit the same behavior.

By using the path integral approach proposed by Ref. \cite{miyaji2015distance}, we are able to compute the QFI without needing to compute the relevant SLD -- an impractical task in a quantum field theoretic context (see Ref. \cite{headley2026path}). 

Yet, let us point that, by recalling the relation $I^{(\psi)}_{ab} = 4g^{(\psi)}_{ab} = 4 \text{Re}\big[G^{(\psi)}_{ab}\big]$ and the definition of $I^{(\psi)}_{ab}$ in Eq. (\ref{eq:Fisher_multiparam_mixed}), the SLD related to the vacuum-state QGT $G_{ab,d+1}$, as given in Eq. (\ref{eq:general_qim}), can still be expressed as
\begin{align}
      L_{a,d+1} = & - 2 \int_{-\infty}^{\infty} d^d \textbf{x} \, \bigg( \int_{-\infty}^{0} d \tau \, \Delta \mathcal{O}_a(x) \, |\Omega_0\rangle  \langle \Omega_0 | \nonumber \\[5pt]
     & + \int_{0}^{\infty} d \tau \, |\Omega_0\rangle  \langle \Omega_0 | \, \Delta \mathcal{O}_a(x) \bigg) ,
\end{align}
where $\Delta \mathcal{O}_a(x) = \mathcal{O}_a(x) - \langle \Omega_0 | \mathcal{O}_a(x) | \Omega_0 \rangle$. This expression includes an ideal measurement, which is not achievable in a QFT setting \cite{sorkin1993impossible}. It also requires integrating the local operator $\Delta \mathcal{O}_a(x)$ over an infinite Euclidean spacetime domain, which is experimentally impractical. Note that successful approaches to QFI calculation have considered system with discrete modes, such as a quantum field in a cavity \cite{ahmadi2014relativistic,ahmadi2014quantum,lindkvist2015motion} or mean-field gauge theory on a lattice \cite{zhou2026quantum}.

Besides the foundational interest of evaluating the scaling of the QFI of bosonic and fermionic field theories, which in a sense quantifies their inherent sensitivity to the mass parameter, there are several possible continuations of this work. In principle, the method used here can be used to falsify any QFT. This can be done in the following way. First, the QFI of a given state with respect to the free parameters of a proposed QFT are calculated. Then, the corresponding QCRLB is computed -- this provides a lower bound on the precision of estimation of the parameters according to the considered QFT. Finally, any observation of the considered state which allows for an estimation of the given parameters at a higher precision (which violates the computed QCRLB) would falsify the QFT. 

Furthermore, this calculation can be performed for any state of the field theory, not just the vacuum state. It would then be worthwhile to contrast the ultimate quantum Cram{\'e}r-Rao bound with specific estimators that can be applied to concrete scenarios, such as the evaluation of Lamb shifts, scattering amplitudes or dipole moments.


\acknowledgments
PA thanks Jack Clarke, Frank Deppisch, and Francis Headley for the valuable discussions. This work was funded by the Leverhulme Trust Research Project Grant RPG-2024-287.

\bibliography{apssamp}

\appendix

\section{Integral evaluation}
\label{subsec:KG_integral}
For some function $F(\cdot)$, consider an integral of the form
\begin{align} 
    & \mathcal{I}_{d+1} = 
    \int^{0}_{-\infty} d\tau \int_{0}^{\infty}d\tau' \int_{-\infty}^{\infty} d^d \textbf{x} \int_{-\infty}^{\infty} d^d \textbf{x}' \, F(|x'-x|) .
\end{align}
To evaluate this integral we perform a coordinate transformation from 
$(\textbf{x},\tau)$ and $(\textbf{x}',\tau')$ to centre-of-mass $(\textbf{X},T)$ and relative $(\textbf{r},t)$ coordinates defined as 
\begin{align} \label{eq:rel_coord_transform}
    \textbf{X} &= \textbf{x}' + \textbf{x} , \quad t = \tau' - \tau  \nonumber \\[5pt]
    \textbf{r} &= \textbf{x}' - \textbf{x} , \quad T = \tau' + \tau.
\end{align}
where $\textbf{X} = ( \, \text{X}_{i} \, | \,  i=1,\dots,d \, )$ and $\textbf{r} = ( \, \text{r}_{i} \, | \,  i=1,\dots,d \, )$. For $i \in \{  1,\dots,d  \}$, the integration volume elements are related as 
\begin{align}
    dx_i \, dx_i' &= |\text{det}\,\textbf{J}_i| d\text{X}_{i} d\text{r}_{i} \nonumber \\[5pt]
    &= \left|\text{det}
    \begin{bmatrix}
    \frac{\partial x_i}{\partial \text{X}_{i}} & \frac{\partial x_i}{\partial \text{r}_{i}} \\[7pt]
    \frac{\partial x'_i}{\partial \text{X}_{i}} & \frac{\partial x'_i}{\partial \text{r}_{i}}
    \end{bmatrix}
    \right| d\text{X}_{i} d\text{r}_{i} \nonumber \\[5pt]
    &= \left| \text{det}
    \begin{bmatrix}
    \frac{1}{2} & -\frac{1}{2} \\[7pt]
    \frac{1}{2} & \frac{1}{2}
    \end{bmatrix}
    \right| d\text{X}_{i} d\text{r}_{i} \nonumber \\[5pt]
    & =\frac{1}{2} d\text{X}_{i} d\text{r}_{i}.
\end{align}
Hence
\begin{align}
    d^d\textbf{x} \, d^d\textbf{x}' &= \frac{1}{2^d}d^d\textbf{X} d^d\textbf{r} \nonumber \\[5pt]
    &= \frac{1}{2^d}d\text{X}_{1} \cdots d\text{X}_{d} \, d\text{r}_{1} \cdots d\text{r}_{d}
\end{align}
and similarly
\begin{equation}
    d\tau d\tau' = \frac{1}{2} dT dt.
\end{equation}
Note that for $\tau \in (-\infty, 0]$ and $\tau' \in [0, \infty)$ we have 
\begin{align}
    T+t &= 2\tau' \in [0, \infty) \nonumber \\[5pt]
    T-t &= 2\tau \in (-\infty, 0],
\end{align}
which implies that $T \in [-t, t]$
for $t \in [0, \infty)$. Therefore, under the coordinate transformation in Eq. (\ref{eq:rel_coord_transform}), $\mathcal{I}_{d+1}$ transforms as
\begin{align} \label{eq:I_1_eq2}
     \mathcal{I}_{d+1} = & \frac{1}{2^{d+1}}
    \int_{0}^{\infty}dt \int^{t}_{-t}dT \int_{-\infty}^{\infty} d^d \textbf{r} \int^{\infty}_{-\infty} d^d \textbf{X} \nonumber \\[5pt] 
    & F \bigg( \sqrt{t^2+\textbf{r}^2} \bigg) \nonumber \\[5pt]
     =& \frac{W_d}{2^{d}}
    \int_{0}^{\infty}dt \int_{-\infty}^{\infty} d^d \textbf{r} \cdot t \cdot F \bigg( \sqrt{t^2+\textbf{r}^2} \bigg) ,
\end{align}
where $W_d \equiv \int_{-\infty}^{\infty} d^d \textbf{X}$.
Let us instead consider the volume
\begin{align}
    V_d &= \int^{\infty}_{-\infty} d^d \textbf{x} = \int^{\infty}_{-\infty} d^d \textbf{x}'
\end{align}
in $(\textbf{x}, \tau)$ or  $(\textbf{x}', \tau')$ coordinates. Note that if $\textbf{x},\textbf{x}' \in [-\bf{k},\bf{k}]$ for $\bf{k} \in \mathbb{R}^{d}_{\geq 0 }$, then $\textbf{X} \in [-2\textbf{k},2\textbf{k}]$. Hence, taking the limit $\textbf{k} \rightarrow \infty$ gives the relation
\begin{align}
    V_{d} &= \lim_{\textbf{k} \rightarrow \infty} \int^{\textbf{k}}_{-\textbf{k}} d^d \textbf{x} 
     = \lim_{\textbf{k} \rightarrow \infty} \frac{1}{2^d} \int^{2\textbf{k}}_{-2\textbf{k}} d^d \textbf{X} 
    = \frac{W_d}{2^d}.
\end{align}
We substitute $V_d$ into Eq. (\ref{eq:I_1_eq2}) and apply a transformation to a hyper-spherical coordinate system $(R, \varphi_{1}, \dots , \varphi_{d})$ in $(d+1)$-dimensional Euclidean space as \cite{blumenson1960derivation}
\begin{align} \label{eq:spherical_coords}
    t &= R \, \cos\big(\varphi_{1}\big) \nonumber \\[5pt]
    \text{r}_{1} &= R \, \sin\big(\varphi_{1}\big) \, \cos\big(\varphi_{2}\big) \nonumber \\[5pt]
    \text{r}_{2} &= R \, \sin\big(\varphi_{1}\big) \, \sin(\varphi_{2}) \, \cos\big(\varphi_{3}\big) \nonumber \\[5pt]
    & \vdotswithin{=} \nonumber \\[5pt]
    \text{r}_{d-1} &= R \, \sin\big(\varphi_{1}\big) \, \cdots \, \sin\big(\varphi_{d-1}\big) \cos\big(\varphi_{d}\big) \nonumber \\[5pt]
    \text{r}_{d} &= R \, \sin\big(\varphi_{1}\big) \, \cdots \, \sin\big(\varphi_{d-1}\big) \sin\big(\varphi_{d}\big).
\end{align}
For $\textbf{r} \in (-\infty, \infty)$ and $t \in [0,\infty)$, the hyper-spherical coordinates in $(d+1)$ dimensions are $R \in [0,\infty)$, $\varphi_{1} \in [0,\frac{\pi}{2}]$, $\varphi_{i} \in [0,\pi]$ for $i = 2, 3, \dots d-1$ and $\varphi_{d} \in [0,2\pi)$ for $d>1$. The Jacobian of this coordinate transformation is \cite{nguyen2014ndimensional}
\begin{align}
    \big|J_{0,d+1}\big| = R^d \, \prod^{d-1}_{k=1} \sin^{d-k} \big(\varphi_{k} \big) .
\end{align}
Hence, if $d > 1$:
\begin{align}
 \mathcal{I}_{d+1} = & V_d \int_{0}^{\frac{\pi}{2}} d\varphi_{1} \,\sin^{d-1}\big(\varphi_{1}\big) \cos\big(\varphi_{1}\big) \nonumber \\[5pt]
    & \prod^{d-1}_{k=2} \, \int_{0}^{\pi} d\varphi_{k} \, \sin^{d-k} \big(\varphi_{k}\big) \int_{0}^{2\pi} d\varphi_{d} \nonumber \\[5pt] 
    &  \int_{0}^{\infty} dR \cdot F \big( R \big)  R^{d+1} \nonumber \\[5pt]
     = & \frac{2 \pi \cdot V_d}{d} \, \prod^{d-1}_{k=2} \bigg( \frac{\sqrt{\pi}  \Gamma \big( \frac{1 + d - k}{2} \big)}{\Gamma \big( \frac{2 + d - k}{2} \big)} \bigg) \nonumber \\[5pt]
    &  \int_{0}^{\infty} dR \, F \big( R \big)  R^{d+1} .
\end{align}
If $d=1$ we substitute $V_d = V_1$ into Eq. (\ref{eq:I_1_eq2}) and transform $(t,\textbf{r}=\text{r})$ to polar coordinates $(R,\varphi)$, where
\begin{align}
    t &= R \, \cos \big(\varphi\big) \nonumber \\[5pt]
    \text{r} &= R \, \sin \big(\varphi\big)
\end{align}
for $R \in [0,\infty)$ and $\varphi \in [-\frac{\pi}{2},\frac{\pi}{2}]$. This gives
\begin{align} \label{eq:I_1_result_d_1_appendix}
 \mathcal{I}_{1+1} = & V_1 \int_{-\frac{\pi}{2}}^{\frac{\pi}{2}} d\varphi \, \cos \big( \varphi \big) \int_{0}^{\infty} dR \cdot F \big( R \big) \cdot R^{2} \nonumber \\[5pt]
 & = 2 V_1 \int_{0}^{\infty} dR \cdot F \big( R \big) \cdot R^{2}.
\end{align}
Finally, in the case where $d=0$ we have the result
\begin{align} \label{eq:I_1_result_d_0_appendix}
    & \mathcal{I}_{0+1} = \int_{0}^{\infty} dR_0  F \big( R_0 \big) \cdot R_0 
\end{align}
for $R_0 = \tau' - \tau$.

\section{Convolution of Klein-Gordon propagators} \label{sec:convolution_klein_gordon_propagators}
Here we evaluate the expression
\begin{equation}
    C = \int_{-\infty}^{\infty}d^{d+1} y \, \Delta(x' - y) \, \Delta(x - y) ,
\end{equation}
where
\begin{align}
    \Delta(x'-x) = & \int_{-\infty}^{\infty}\frac{d^{d+1}p}{(2\pi)^{d+1}} \frac{e^{-ip\cdot(x'-x)}}{p^2 + m^2} \nonumber \\[5pt]
    = & \frac{m^{\frac{d-1}{2}}}{(2 \pi)^{\frac{d+1}{2}}}|x'-x|^{\frac{1-d}{2}} K_{\frac{d-1}{2}} \big( m|x'-x| \big)
\end{align}
is the Klein-Gordon (KG) propagator in $(d+1)$-dimensional Euclidean spacetime. Consider the Fourier transform $\tilde{\Delta}(p)$ of $\Delta(x'-x)$, defined by
\begin{equation} \label{eq:kg_prop_fourier_transf}
    \Delta(x'-x) = \int_{-\infty}^{\infty}\frac{d^{d+1}p}{(2\pi)^{d+1}}e^{-ip\cdot(x'-x)}\tilde{\Delta}(p),
\end{equation}
where $p=(\textbf{p},p^0)$ is a $(d+1)$-dimensional momentum vector, and
\begin{align} \label{eq:delta_p_def}
     \tilde{\Delta}(p)=\frac{1}{p^2 + m^2}.
\end{align}
Substituting Eq. (\ref{eq:kg_prop_fourier_transf}) into the convolution $C$ gives
\begin{align}
    C  = & \int_{-\infty}^{\infty}d^{d+1} y \, \int_{-\infty}^{\infty}\frac{d^{d+1}k}{(2\pi)^4} \int_{-\infty}^{\infty}\frac{d^{d+1}p}{(2\pi)^{d+1}} e^{-i y \cdot (k - p)}  \nonumber \\[5pt]
    &   e^{-i p \cdot x' + i k \cdot x} \tilde{\Delta}(p) \, \tilde{\Delta}(k).
\end{align}
Using the integral representation of the delta function, $C$ simplifies to
\begin{align}
    C = & \int_{-\infty}^{\infty}d^{d+1}k \int_{-\infty}^{\infty}\frac{d^{d+1}p}{(2\pi)^{d+1}} \, \delta(k - p) \, e^{-i p \cdot x' + i k \cdot x} \nonumber \\[5pt]
    &  \tilde{\Delta}(p) \, \tilde{\Delta}(k) \nonumber \\[5pt]
    = & \int_{-\infty}^{\infty}\frac{d^{d+1}p}{(2\pi)^{d+1}} \, e^{-i p \cdot (x - x')} \, \tilde{\Delta}^2(p)  \nonumber \\[5pt]
    = & \int_{-\infty}^{\infty}\frac{d^{d+1}p}{(2\pi)^{d+1}} \, e^{-i p \cdot (x - x')} \, \frac{1}{(p^2 + m^2)^2} ,
\end{align}
where we substituted Eq. (\ref{eq:delta_p_def}). Next, we use Schwinger parametrization to write
\begin{equation}
    \frac{1}{(p^2 + m^2)^2} = \int^{\infty} da \cdot a \, e^{-a(p^2 + m^2)} .
\end{equation}
Hence
\begin{align}
    C & = \int^{\infty} da \int_{-\infty}^{\infty}\frac{d^{d+1}p}{(2\pi)^{d+1}} \cdot a \, e^{-a m^2} \, e^{-ap^2 -i p \cdot (x - x')} .
\end{align}
By completing the square, 
we simplify $C$ to
\begin{align}
    C  = & \int^{\infty} da \int_{-\infty}^{\infty}\frac{d^{d+1}p}{(2\pi)^{d+1}} a \, e^{-a m^2} \, e^{-\frac{(x'-x)^2}{4a}} \nonumber \\[5pt]
    & e^{-a \big(p + \frac{i (x'-x)}{2a}\big)^2} \nonumber \\[5pt]
     = & \frac{1}{(2\pi)^{d+1}} \int^{\infty} da a \, e^{-a m^2} \, e^{-\frac{(x'-x)^2}{4a}} \bigg( \frac{\pi}{a} \bigg)^{\tfrac{d+1}{2}} \nonumber \\[5pt]
    = & \frac{1}{2(2\pi)^{\tfrac{d+1}{2}}} \bigg( \frac{m}{|x'-x|} \bigg)^{\frac{d-3}{2}} K_{\frac{d-3}{2}} (m |x'-x|) .
\end{align}

\section{Dirac propagator} \label{sec:dirac_propagator}
Here, for completeness, we report a derivation of the position-space expression of the Dirac propagator $\Delta_D(x'-x)$  in $(d+1)$-dimensional Euclidean spacetime. In $(d+1)$-dimensional Minkowski spacetime the following relation holds between the Dirac propagator $\Delta_D(x'-x)$ and the KG propagator $\Delta(x'-x)$ \cite{peskin1995introduction}
\begin{align}
    \Delta_D(x'-x) =(i \slashed{\partial}'_x + m)\Delta(x'-x) ,
\end{align}
where we use the Feynman slash notation $\slashed{\partial}'_x = \gamma \cdot \partial'_x$ of the gradient $\partial'_x$ with respect to $x'$. By performing Wick rotation as
\begin{align} \label{eq:Wick_rot_gamma}
    & t \to \tau = it = x_0, \quad \vec{x} \to \vec{x}, \nonumber \\[4pt]
    & {\gamma}^0 \to {\gamma}^0, \quad \vec{\gamma} \to i\vec{\gamma},
\end{align}
we obtain $\Delta_D(x'-x)$ as
\begin{align}
 \Delta_D(x'-x) = & (\slashed{\partial}'_x + m)\Delta(x'-x) \nonumber \\[5pt]
    = & (\slashed{\partial}'_x + m) \frac{m^{\frac{d-1}{2}}}{(2 \pi)^{\frac{d+1}{2}}}|x'-x|^{\frac{1-d}{2}} K_{\frac{d-1}{2}} \big( m|x'-x| \big)  \nonumber \\[5pt]
    = & \frac{m^{\frac{d-1}{2}}}{(2 \pi)^{\frac{d+1}{2}}} \gamma \cdot \partial'_x \frac{K_{\frac{d-1}{2}} \big( m|x'-x| \big)}{|x'-x|^{\frac{d-1}{2}}} \nonumber \\[5pt]
    & + \frac{m^{\frac{d+1}{2}}}{(2 \pi)^{\frac{d+1}{2}}} \cdot \frac{K_{\frac{d-1}{2}} \big( m|x'-x| \big)}{|x'-x|^{\frac{d-1}{2}}} ,
\end{align}
where we substituted the KG propagator in Euclidean spacetime, as given in Eq. (\ref{eq:kg_propagator_def}). In order to simplify this expression, we evaluate the term
\begin{align}
    & \gamma \cdot \partial'_x \frac{K_{\frac{d-1}{2}} \big( m|x'-x| \big)}{|x'-x|^{\frac{d-1}{2}}} = \sum_{i=0}^{d} \gamma^i  \partial'_i \frac{K_{\frac{d-1}{2}} \big( m|x'-x| \big)}{|x'-x|^{\frac{d-1}{2}}}.
\end{align}
The derivative with respect to $x'_i$ is
\begin{align}
    \partial'_i \frac{K_{\frac{d-1}{2}} \big( m|x'-x| \big)}{|x'-x|^{\frac{d-1}{2}}} = - m (x'_i - x_i) \frac{K_{\frac{d+1}{2}} \big( m|x'-x| \big)}{|x'-x|^{\frac{d+1}{2}}} .
\end{align}
Hence, summing all terms together gives
\begin{align}
    & \gamma \cdot \partial'_x \frac{K_{\frac{d-1}{2}} \big( m|x'-x| \big)}{|x'-x|^{\frac{d-1}{2}}} \nonumber \\[5pt]
    & = - \bigg( \sum_{i=0}^{d}\gamma^i(x'_i-x_i) \bigg) \frac{m K_{\frac{d+1}{2}} \big( m|x'-x| \big)}{|x'-x|^{\frac{d+1}{2}}} .
\end{align}
Therefore, the Dirac propagator in $(d+1)$-dimensional Euclidean spacetime is
\begin{align}
    \Delta_D(x'-x) = & \left(\frac{m}{2\pi |x'-x|}\right)^{\frac{d+1}{2}} \bigg[ |x'-x| K_{\tfrac{d-1}{2}}(m|x'-x|) \nonumber \\[5pt]
    & - \sum_{i=0}^{d}\gamma^i(x_i'-x_i) \, K_{\tfrac{d+1}{2}}(m|x'-x|) \bigg] .
\end{align}

\end{document}